\documentclass[journal,10pt]{IEEEtran}
\ifCLASSINFOpdf
\else
\fi
\usepackage{enumitem}
\usepackage[noend]{algpseudocode}
\usepackage{mathtools}
\usepackage{bbm}
\usepackage{algorithm,tabularx}
\usepackage{setspace}
\usepackage{graphicx}
\usepackage{notoccite}
\usepackage{amssymb,balance}
\usepackage{url}
\usepackage[table]{xcolor}
\usepackage{placeins}
\usepackage{breqn}
\usepackage{amsfonts}
\usepackage{amsthm}
\usepackage{amsmath}
\usepackage{bbm}
\usepackage{subfigure}
\usepackage{multirow}
\usepackage{pdfrender}
\usepackage{color}
\usepackage{accents}

\makeatletter
\newcommand{\multiline}[1]{%
  \begin{tabularx}{\dimexpr\linewidth-\ALG@thistlm}[t]{@{}X@{}}
    #1
  \end{tabularx}
}
\makeatother

\begin{document}
\sloppy
\title{On the Downlink Performance of RSMA-based UAV Communications}

\author{Wael Jaafar,~\IEEEmembership{Senior Member,~IEEE}, Shimaa~Naser,
        Sami~Muhaidat,~\IEEEmembership{Senior~Member,~IEEE},\\     Paschalis C. Sofotasios,~\IEEEmembership{Senior~Member,~IEEE} and Halim Yanikomeroglu,~\IEEEmembership{Fellow,~IEEE}

\thanks{Copyright (c) 2015 IEEE. Personal use of this material is permitted. However, permission to use this material for any other purposes must be obtained from the IEEE by sending a request to pubs-permissions@ieee.org.}

\thanks{W. Jaafar and H. Yanikomeroglu are with the Department of Systems and Computer Engineering, Carleton University, Ottawa,
ON, Canada, (e-mails: \{waeljaafar, halim\}@sce.carleton.ca.)}
\thanks{S. Naser and S. Muhaidat are with the Department of Electrical and Computer Engineering, Khalifa University, Abu Dhabi, UAE, (e-mails:\{100049402, sami.muhaidat\}@ku.ac.ae.)}
\thanks{P. C. Sofotasios is with the Department of Electrical Engineering and Computer Science, Khalifa
University of Science and Technology, Abu Dhabi, UAE, and also with the Department of Electrical Engineering, Tampere University, Tampere, Finland, (email: p.sofotasios@ieee.org.)}
\thanks{This work is supported in part by the Natural Sciences and Engineering Research Council Canada (NSERC) and in part by Khalifa University under Grant
No. KU/RC1-C2PS-T2/8474000137 and Grant No. KU/FSU-474000122.}
}

\markboth{IEEE Transactions on Vehicular Technology,~Vol.~XX, No.~XX, Month~Year}%
{Shell \MakeLowercase{\textit{et al.}}: Bare Demo of IEEEtran.cls for IEEE Journals}

\maketitle

\begin{abstract}
The use of unmanned aerial vehicles (UAVs) as base stations (BSs) is envisaged as a key enabler for {the fifth generation (5G)} and beyond-5G networks. Specifically, aerial base stations (UAV-BS) are expected to provide ubiquitous connectivity and high spectral efficiency. To this end, we present in this correspondence {an} in-depth look into the integration of rate-splitting multiple access (RSMA) with UAV-BSs and downlink transmissions. A {non-convex} problem of {joint} UAV placement, RSMA precoding, and rate splitting, aiming to maximize the weighted sum data rate of users is formulated. {Due to its complexity, two sub-problems are investigated, namely the UAV placement and RSMA parameters optimization. The resulting solutions are then combined to propose a novel alternating optimization method}. Simulation results illustrate {the latter's} efficiency compared to baseline approaches.
%a sub-optimal approach is proposed where the UAV location is first optimized, and then a solution to the joint beamforming and rate-splitting problem is found. 
%Simulation results illustrate the efficiency and robustness of RSMA compared to other MA techniques. 

%Finally, the impact of some parameters, e.g., number of UAV antennas and users’ locations, is further investigated.

%If combined with an efficient multiple access (MA) technique, it is expected that ground users performances are significantly improved. In this matter, rate-splitting multiple access (RSMA) is seen as a promising technique capable of supporting large-scale systems with different environment conditions. Hence, we present in this letter the first in-depth look into the integration of RSMA to the UAV-BS downlink. The non-covex problem of UAV placement and joint RSMA beamforming and rate-splitting, aiming at maximizing the weighted sum data rate of users, is formulated. Then, a sub-optimal approach is proposed. The UAV location is optimized at first, followed by the joint beamforming and rate-splitting problem solving. The simulation results illustrate the efficiency and robustness of RSMA compared to other MA techniques. Finally, the impact of some parameters, e.g., number of UAV antennas and users' locations, is investigated.    
\end{abstract}


\begin{IEEEkeywords}
MIMO, rate-splitting multiple access, RSMA, unmanned aerial vehicle.
\end{IEEEkeywords}

\IEEEpeerreviewmaketitle

%\vspace{-10pt}
\section{Introduction}
There has been a  growing interest in unmanned aerial vehicle (UAV) communications, seen as a promising enabler for the {fifth generation (5G) and beyond-5G networks} \cite{Bor2019}--\cite{Li2019}. The deployment of UAVs, driven by the emergence of new applications, such as aerial security inspection, smart agriculture, and aerial delivery, is expected to continue shaping future breakthrough services. UAVs can also serve as aerial base stations (UAV-BS) in order to support ubiquitous connectivity, particularly in rural areas and disaster areas, and high data rates, when deployed in dense areas. In this case, the UAV-BS will be a source of/or experience interference as in conventional cellular networks. Fortunately, such a problem can be addressed by the additional degrees of freedom provided solely by a UAV, i.e., the design of {three-dimensional (3D)} location and high probability of line-of-sight (LoS) to ground users \cite{Alzenad2018}. Nevertheless, these additional degrees of freedom are not sufficient to provide the best experience to a constantly growing number of ground users. In this respect, downlink multiple access techniques play an important role in realizing the data rate requirements, low latency, and connectivity, without added resources.   

{Recently, rate-splitting multiple access (RSMA) has been identified as a highly-reliable and spectrum-efficient multiple access scheme, that is capable of outperforming both non-orthogonal multiple access (NOMA) and space division multiple access (SDMA) \cite{Mao2018}. RSMA relies on the implementation of a linear precoder at the transmitter and successive interference cancellation (SIC) at the receiver. The process starts by dividing user's messages into common and private parts at the transmitter. The common parts of (all and/or subsets of) users are combined together and encoded into a single common stream, while the private parts are encoded into distinct private streams. These streams are superimposed and sent over a multiple-input multiple-output (MIMO) channel. Then, each user decodes the first common stream and recovers its own data. At the receiver, the interference from the decoded common stream is removed using the SIC. This is followed by successive decoding of the next common parts (of users' subsets) and removing them, then by private part decoding, while treating other (common and) private signals as noise.}

{In \cite{Mao2018}, Mao \textit{et al.} showed that RSMA outperforms NOMA and SDMA systems over a wide range of network loads and user deployment scenarios, and that RSMA has a lower-complex receiver design than NOMA. In \cite{Mao_conf2018}--\nocite{Mao2018_2}\cite{Mao2018_3}, it was shown that RSMA provides better energy and spectral efficiency than NOMA and SDMA for several user deployments, and unicast and multicast transmissions. Extensions of these works have been made to other system models, e.g., downlink coordinated multi-point transmission \cite{Mao_conf2019}, cloud-radio access network (C-RAN) \cite{Yu_2019}, multi-user multi-antenna wireless information and power transfer (SWIPT) \cite{Mao2019spawc}, hybrid radar-communication system \cite{xu2020ratesplitting}, and multi-beam satellite communications \cite{yin2020ratesplitting}. 
 {In \cite{Rahmati2019}, downlink RSMA was considered from one ground BS to serve several aerial users only, whereas in} \cite{Ahmad2019}, a UAV-assisted C-RAN system, where RSMA parameters were optimized for joint
transmissions from ground and aerial BSs to terrestrial users, was proposed. %and RSMA configuration (i.e., one common signal per user). 
%For UAV systems, \cite{Ahmad2019} proposed UAV-assisted C-RAN where RSMA parameters were optimized for joint transmission from ground BSs and UAV-BSs to users, while \cite{Rahmati2019} used RSMA at the ground BS to serve aerial users. 
 %However, \cite{Rahmati2019} did not investigate the UAV-BS case, while \cite{Ahmad2019} ignored UAV-BS placement optimization and adopted simplified air-to-ground channel model and RSMA configuration (one common signal per user).
}
{UAV-BS placement has been previously considered with multiple access, thus the UAV-BS' altitude and its horizontal location can be either separately or jointly optimized. In \cite{Yin2019}, the authors investigated joint UAV-BS placement, power and time duration allocations for time-division and frequency-division multiple access (TDMA and FDMA). Xia \textit{et al.} found in \cite{Xiao2016} optimal SDMA user grouping and precoding for ground-to-air (user to UAV-BS) communications in order to maximize the achieved sum data rate.
Also, UAV-BS placement and NOMA power allocation were separately solved in \cite{Liu2019} to maximize the system's sum rate, whereas an extension to joint NOMA power allocation, user pairing, and UAV-BS placement has been studied in \cite{Nguyen2019_inrs}.} 

{However, the aforementioned works either focused on simple and well-investigated multiple access techniques such as TDMA, FDMA, SDMA, and single-input-single-output (SISO) NOMA, which are not flexible, and are typically designed for specific scenarios (e.g., single-antenna BS, small number of users, etc.) \cite{Yin2019}--\cite{Nguyen2019_inrs}, or ignored UAV-BS placement optimization when using RSMA on the air-to-ground channels \cite{Ahmad2019}. It is obvious that given the recent research interest into RSMA, the latter has not been sufficiently researched in the context of aerial communications, which motivates us to combine the UAV and RSMA technologies to improve wireless data transmissions to ground users. Specifically, we investigate the joint optimization of UAV placement, RSMA precoding, and rate splitting, in order to maximize the weighted sum rate (WSR) of the downlink communication. Since the optimization problem is non-convex, we propose an alternating optimization (AO) method, where the main problem is divided into two sub-problems, namely UAV placement and RSMA parameters optimization,  respectively solved for fixed RSMA configuration (using successive convex optimization) and UAV location (using the weighted minimum mean square error method), then combined into a global algorithm, where the optimization approaches for the sub-problems are iteratively executed.}

%Finally, the impact of some key parameters (e.g., number of transmit antennas of UAV-BSs and user locations) on the system performance is studied.

%Since the optimization problem is non-convex, we proposed a sub-optimal iterative searching method, where in each iteration, the UAV location is fixed and the RSMA parameters are optimized using the weighted minimum mean square error (WMMSE) method. Finally, the impact of some key parameters (e.g., number of transmit antennas of UAV-BSs and user locations) on system performance is studied.

%\textcolor{red}{Since the optimization problem is non-convex, it has been separated into two sub-problems. In the first, the UAV location is optimized for a long-term placement, while in the second, the joint beamforming and rate-splitting problem is solved using the weighted minimum mean square error (WMMSE) method.} Finally, the impact of some key parameters (e.g., number of transmit antennas of UAV-BSs and user locations) on system performance is studied. 

The rest of the paper is organized as follows. In Section II, the system model is presented and the problem is formulated. Section III details the proposed solutions {and discusses their complexity}, while Section IV presents the results. Finally, Section V concludes the paper.   

%\begin{figure}[t]
%	% Requires \usepackage{graphicx}
%	\centering
%	\includegraphics[width=200pt]{Fig_model.pdf}
%	\caption{System model.}
%	\label{Fig:model}
%\end{figure}

\textit{Notations:} In the remainder of the paper, boldface uppercase and lowercase represent matrices and vectors, respectively. $(\cdot )^{T}$ and $(\cdot )^{H}$ denote the transpose and Hermitian operations respectively. $\mathbb{E}(\cdot)$ is the expectation, {$\Re(\cdot)$ is the real part of a complex number}, $||\cdot||$ is the Euclidean norm, $|\cdot|$ is the absolute value, \textbf{\textit{0}} is the zero matrix, {\textbf{\textit{I}} is the identity matrix}, {$\mathbbm{1}_{a \times b}$ is $a \times b$ all-ones matrix, and} tr$(\cdot)$ is the trace of a matrix.
%, and $L_{\frac{1}{2}}(x)={}_1F_1(-\frac{1}{2};1;x)$ is the Laguerre polynomial, defined using the confluent hypergeometric function of the first kind ${}_1F_1$.

%\vspace{-15pt}
\section{System Model and Problem Formulation}

%and $\mathcal{N}(0,\sigma^2)$ is a real-value Gaussian distribution with zero mean and variance $\sigma^{2}$. Let $\textbf{z}=[z_1,\ldots,z_Z]$ be a vector of length $Z$, then $L_1(\textbf{z})=\sum_{i=1}^Z |z_i|$ is the $L_1$ norm.

\subsection{System Model}
{We consider a downlink one layer rate-splitting (RS) transmission where a hovering UAV-BS, equipped with $N_t$ antennas, is communicating with $K$ single-antenna users, using a dedicated frequency band. At the UAV side, the distinct messages of all users denoted by $M_k$ ($k=1,2,...,K$) are initially divided into common and private parts as $M_k=\{M_k^{p}, M_k^{c} \}$. Then, all common messages from all users are encoded together into a single stream $s_{0}$, for the purpose of reducing interference. This stream $s_0$ will be eventually decoded by all users. Meanwhile, the private message of each user is encoded into a separate private stream $s_k$ ($k=1,2,...,K$) that will be decoded by the corresponding user only. Hence, the vector of the $K+1$ streams to be transmitted is denoted by $\textbf{s}=[s_{0},s_{1},s_{2},...,s_K]^T$.
It is to be noted that a generalized RS model can be obtained by following the same steps as in \cite{Mao2018}. For the sake of simplicity, we keep such design out of the scope of this paper, and focus on the simplified version of one common signal only. In order to reduce multi-user interference, precoding $\textbf{P}=[\textbf{p}_{0},\textbf{p}_{1},\textbf{p}_{2},...,\textbf{p}_{K}]_{N_t \times (K+1)}$ is utilized to give the streams appropriate weights at each transmitting antenna. Finally, signals are superimposed and broadcast as $\textbf{x}=[x_1,\ldots,x_{N_t}]^T=\textbf{P}\;\textbf{s}$.}  
Hence, the received signal at user $k$ can be written as
\begin{equation}
\small
\label{eq:receiver}
    y_k=\mathbf{h}_k^H \mathbf{P} \mathbf{s} + n_k,\qquad \forall k=\{1,\ldots,K\},
\end{equation}
where $n_k$ is the circularly symmetric complex additive white Gaussian noise (AWGN) with zero-mean and variance $\sigma^2$, and $\mathbf{h}_k=[h_k^1,\ldots, h_k^{N_t}]^T \in {\mathbb{R}^{N_t \times 1}}$ is the air-to-ground channel from the UAV to user $k$. 
{We assume that the air-to-ground communication channels are dominated by line-of-sight (LoS) links, thus,} the channel coefficient $h_k^j$ ($j=1, \ldots, N_t$) follows {the free-space path loss model}
%Rician distribution with parameter \textcolor{black}{$\alpha(\theta_k)$ \cite{Azari2018}} and is
expressed by {\cite{Zeng2016}}
\begin{equation}
\small
    \label{eq:Ricechannel}
    h_k^{j}={{d_k}^{-{{\frac{\beta}{2}}}}}
    %\textcolor{black}{g_k^j(\alpha(\theta_k))}
    , \qquad \forall j =\{1,\ldots,N_t\},
    %h_k^{j}={d_k^{-{\textcolor{black}{\beta(\theta_k)}}}}\left(\sqrt{\frac{\textcolor{blue}{\alpha(\theta_k)}}{1+\textcolor{blue}{\alpha(\theta_k)}}}\bar{h}_k^j+\sqrt{\frac{1}{1+\textcolor{blue}{\alpha(\theta_k)}}} \tilde{h}_k^j\right),\; \forall j %=1,\ldots,N_t, \nonumber
\end{equation}
where $d_k=||\textbf{q}-\textbf{q}_k||_2$ is the distance between the UAV and user $k$, $\textbf{q}=[x,y,z]$ and $\textbf{q}_k=[x_k,y_k,z_k]$ are the UAV and user $k$ 3D locations, respectively, {and $\beta=2$ is the free space path-loss factor \cite{Zeng2016}.}
%\footnote{\textcolor{black}{For simplicity, the UAV's altitude $z$ is assumed fixed, and that users are on the ground, i.e. $z_k=0$, $\forall k$.}}
 %\textcolor{blue}{and $\theta_k=\arcsin{\left({z}/{d_k}\right)}$} is the elevation angle between user $k$ and the UAV. %\textcolor{black}{Also, $g_k^j(\alpha(\theta_k))$ is the small-scale fading, which follows a Rician distribution with K-factor $\alpha(\theta_k)$ and $\mathbb{E}\left\{ ||g_k^j||^2 \right\}=1$. 
%The Rician model is adequate for air-to-ground channels due to the possible combination of LoS and NLoS multipath scatterers experienced at users.}
%$\bar{h}_k^j$ is the deterministic LoS component with $|\bar{h}_k^j|=1$, and $\tilde{h}_k^j$ is the random NLoS component, modeled by a complex Gaussian distribution $\mathcal{CN}(0,1)$. 
%\textcolor{black}{Also, 
%$\alpha(\theta_k)=a_1 e^{\theta_k b_1}$, such that $a_1=\alpha(0)$ and $b_1=\frac{2}{\pi}\text{ln}\left(\frac{\alpha(\frac{\pi}{2})}{\alpha(0)} \right)$ are constants that depend on the environment and system parameters, while 
%\textcolor{blue}{$\beta(\theta_k)=a_1 Pr_{\rm LoS}(\theta_k)+b_1$} is the path-loss factor, where $Pr_{\rm LoS}(\theta_k)$ is the LoS probability, expressed by
%\textcolor{blue}{
%\begin{equation}
%\small
%    \label{PLoS}
%    Pr_{\rm LoS}(\theta_k)=\frac{1}{\left( 1 + a_2 e^{- b_2 (\theta_k- a_2)} \right)},\qquad \forall k \in \{1,\ldots, K\},
%\end{equation}
%and $a_i$ and $b_i$ ($i=1,2$) are constants \cite{Azari2018,Bor2016}.}}
{Following \cite{Mao2018_3,Ahmad2019}, channels are assumed perfectly known at the transmitter and receivers.}
%\footnote{\textcolor{black}{Practically, a quantized/corrupted channel information may be available due to several issues, e.g., UAV mobility and outdated feedback \cite{Dai2019}.}}.} 
Also, we assume that {$\mathbb{E}\{\textbf{s} \textbf{s}^H\}=\textbf{\textit{I}}$} and that the transmit power budget at the UAV is constrained by $\text{tr}(\textbf{P}\textbf{P}^H)\leq P_t$, where $P_t$ is the UAV's maximal transmit power.

%Let $C=\{x,y,h\}$ be the 3D hovering location of the UAV. Hence, the 

{
The decoding procedure at the $k^{th}$ user is described as follows. Since the  common stream $s_c$ is allocated the highest power, it will be decoded first by treating the rest of the received signal as noise. Then, the $k^{th}$ user will extract its intended information from the common stream. In order to improve the detection of the private stream, each user will apply SIC to remove the effect of the common stream. Finally, the $k^{th}$ user will decode its intended private stream $s_k$ by considering the rest of the other users' private streams as noise. Hence, the received {signal-to-interference-plus-noise ratios (SINRs)} of the common and private streams $s_{0}$ and $s_k$ at $k^{th}$ user, can be given by}
{
\begin{equation}
\small
    \label{eq:SINR_c}
    \gamma_{k}^{c}=\frac{|\mathbf{h}_k^H \mathbf{p}_{0}|^2}{\sum \limits_{i=1}^K |\mathbf{h}_k^H \mathbf{p}_i|^2+\sigma^2}=\frac{d_k^{-2}|\mathbbm{1} \mathbf{p}_0|^2}{\sum \limits_{i=1}^K d_k^{-2}|\mathbbm{1} \mathbf{p}_i|^2|+\sigma^2}, \; \forall k \in \{1,\ldots,K\}
\end{equation}
and
\begin{equation}
\small
    \label{eq:SINRk}
    \gamma_k^{{k}}=\frac{|\mathbf{h}_k^H \mathbf{p}_{k}|^2}{\sum \limits_{\substack{i=1\\i \neq k}}^K |\mathbf{h}_k^H \mathbf{p}_i|^2+\sigma^2}=\frac{d_k^{-2}|\mathbbm{1} \mathbf{p}_k|^2}{\sum \limits_{\substack{i=1\\ i \neq k}}^K d_k^{-2}|\mathbbm{1} \mathbf{p}_i|^2|+\sigma^2}, \; \forall k \in \{1,\ldots,K\},
\end{equation}
where $\mathbbm{1}=\mathbbm{1}_{1 \times N_t}$.}
Let $R_k^j=B \; \text{log}_2\left(1+\gamma_k^j\right)$ be the data rate corresponding to the received SINR $\gamma_k^j$, $\forall k\in \{1,2,...,K\}$, $\forall j\in \{c,1, 2,...,K\}$, where $B$ is the used bandwidth.
Then, in order to ensure that the common message {$s_{0}$} is successfully decoded by all users, the achievable common data rate should not exceed $R_{c}=\text{min}\left\{R_{1}^{c},R_2^{c},...,R_{K}^{c}\right\}$.
We denote {by} $R_{k,\rm{com}}$ the portion of the common rate allocated to user $k$. Then, we have 
{$R_{c}=\sum \limits_{k=1}^K R_{k,\rm{com}}$}.
Finally, the overall achievable data rate of user $k$ can be expressed as 
\begin{equation}
R_{k,\rm{ov}}=R_{k,\rm{com}}+R_k^{{k}}. \; \; \forall k \in \{1,...,K\}.
\end{equation}

%$R_{k,\rm{ov}}=R_{k,\rm{com}}^{k\bar{k}\ubar{k}}+R_{k,\rm{com}}^{k \bar{k}}+R_{k,\rm{com}}^{k \ubar{k}}+R_k^{\textcolor{blue}{k}}$.

%, where $R_k=B \; \text{log}_2\left(1+\gamma_k^{\textcolor{blue}{k}}\right)$, $\forall (k,\bar{k},\ubar{k}) \in \{(1,2,3),(2,3,1),(3,1,2)\}$.

 %For any subset of users $\mathcal{U}\subseteq \mathcal{K}$, the UAV transmits the data stream $s_{\mathcal{U}}$ that can be decoded by these users, while treated as noise by the others. This data stream results from the combined encoding of the splitted messages of users in subset $\mathcal{U}$. Indeed, user $k \in \mathcal{K}$ message is split as $\{M_k^{\mathcal{U}'}|\; \mathcal{U}' \subseteq \mathcal{K}, k \in \mathcal{U}'\}$, and messages of users with the same superscript $\mathcal{U}$, i.e., $\{M_{k'}^{\mathcal{U}}|\; k' \in \mathcal{U} \}$, are encoded into stream $s_{\mathcal{U}}$. The 

%\vspace{-5pt}
\subsection{Problem Formulation}
In this paper, we jointly optimize the {precoding} matrix $\textbf{P}$, the common rate vector $\textbf{r}=[R_{1,\rm{com}},R_{2,\rm{com}},...,R_{K,\rm{com}}]$, and the UAV {3D} location, with {the aim of} maximizing the weighted sum of overall achievable data rates (WSR), defined as $R_{\rm{ov}}=\sum_{k=1}^K w_k R_{k,\rm{ov}}$, where $w_k$ ($k=1,\ldots,K$) is the weight reflecting $k$th user {traffic} priority\footnote{{For simplicity, we assume that each user is demanding a specific service with a given priority. If the priority is the same for all traffic demands, then WSR becomes the sum data rate ($\forall w_k=1$) or average data rate ($\forall w_k=\frac{1}{K}$).}}. For a given weight vector $\textbf{w}=[w_1,\ldots,w_K]$, 
the optimization problem can be formulated as follows:

\begin{subequations}
\small
	\begin{align}
	\max_{\mathbf{P},\mathbf{r},\mathbf{q}} & \quad 
	{\text{WSR}=R_{\rm{ov}}=\sum_{k=1}^K w_k R_{k,\rm{ov}}} \tag{P1} \\
	\label{c1}
	\text{s.t.}\quad & \sum_{k=1}^K R_{k,\rm{com}}\leq R_{c} \tag{P1.a} \\
	\label{c2_1} & R_{k,\rm{ov}}\geq R_{k,\rm{th}},\;\forall k \in \{1,\ldots,K\} \tag{P1.b} \\
	\label{c3} & \text{tr}(\mathbf{P}\mathbf{P}^H)\leq P_t  \tag{P1.c}\\
		\label{c4} & \mathbf{r}\geq \mathbf{0}  \tag{P1.d}\\
	\label{c5} & {x_{\rm{min}}\leq x \leq x_{\rm{max}}, y_{\rm{min}}\leq y \leq y_{\rm{max}}, z_{\rm{min}}\leq z \leq z_{\rm{max}}},  \tag{P1.e}
	\end{align}
\end{subequations}
where 
%$x_{\rm{min}}$ and $x_{\rm{max}}$, and $y_{\rm{min}}$ and $y_{\rm{max}}$, are the UAV's x-axis \textcolor{blue}{and} y-axis displacement limits, respectively\footnote{\textcolor{blue}{We assume here that the UAV's altitude is fixed to $z$.}}. 
$R_{k,\rm{th}}$ is the minimum required rate at user $k$ to ensure respect of QoS, {and ($x_{\min}$,$x_{\max},y_{\min}$,$y_{\max}$,$z_{\min}$,$z_{\max}$) are the minimum and maximum 3D placement coordinates.} %Constraints (\ref{c1})-(\ref{c2}) guarantee that common rate allocations respect the achievable common data rate. 
Problem (P1) is highly non-convex. In fact, for a given UAV location, the problem reduces to WSR maximization by optimizing the precoding matrix and rate-splitting for one layer RSMA, as in \cite{Joudeh2016}. It has been shown that the latter problem is non-convex and non-trivial due to the appearance of the precoder weights in the denominator of the SINR equations \cite{Christensen2008}. Thus, by reduction, we deduce that (P1) is non-convex.  

\section{Proposed {Solution}}
In order to solve the joint problem in (P1), {we propose a low-complexity iterative algorithm, based on {the alternate optimization of the UAV placement and RSMA parameters.} % alternating optimization (AO). 
%To this end, RSMA beamforming and rate splitting are iteratively optimized for several potential UAV locations, until the best solution is obtained.
}

%with the UAV's 3D location in an alternating manner, until convergence.}

%In order to solve (P1), we take the following approach. First, we focus on the UAV placement optimization. Then, we solve the joint beamforming and rate-splitting problem.

\subsection{UAV Placement Optimization}
%\textcolor{blue}{For the sake of simplicity, we assume that the path-loss coefficient $\beta(\theta_k)$ for any channel is fixed to $\beta$.}\footnote{\textcolor{blue}{The objective of this assumption is to reduce the resolution complexity of the UAV placement problem.}} 
{Assuming} that $\textbf{P}$ and $\textbf{r}$ are given, then (P1) reduces to the following UAV placement problem (P2):

\begin{subequations}
\small
	\begin{align}
	\max_{\mathbf{q},\boldsymbol{\eta}} & \quad {
	{\sum_{k=1}^K w_k \; \eta_k } } \tag{P2} 
	\label{c2_21}\\
		\label{c2_0} & 
	{\eta_k \leq \text{log}_2 \left(1+{|\mathbbm{1} \mathbf{p}_k|^2}/\left({\sum \limits_{\substack{i=1\\i\neq k }}^K |\mathbbm{1} \mathbf{p}_i|^2 + \sigma^2 ||{\mathbf{q}-\mathbf{q}_k}||_2^{2}}\right)  \right), \; \forall k,} \tag{P2.a}
	\end{align}
\end{subequations}
\begin{subequations}
\small
	\begin{align}
	\label{c2_2} & {||{\mathbf{q}-\mathbf{q}_k}||_2^2 \leq {\frac{1}{\sigma^2} \left( \frac{|\mathbbm{1}\mathbf{p}_k|^2}{\Lambda_k}-\sum_{\substack{i=1\\ i \neq k}}^K|\mathbbm{1} \mathbf{p}_i|^2 \right), \; \forall k}, \tag{P2.b}}\\
	\label{c2_4} & {x_{\rm{min}}\leq x \leq x_{\rm{max}}, y_{\rm{min}}\leq y \leq y_{\rm{max}}, z_{\rm{min}}\leq z \leq z_{\rm{max}}, \tag{P2.c}}
	\end{align}
\end{subequations}
where {$\eta_k \geq 0$ is a SINR slack variable, which respects constraint (\ref{c2_0}), $\Lambda_k=2^{A_k/B}-1$, $A_k=R_{k,\rm{th}}- R_{k,\rm{com}}$, and constraint (\ref{c2_2}) is equivalent to (\ref{c2_1}). 
Problem (P2) is non-convex due to the non-convexity of constraint (\ref{c2_0}) with respect to $\textbf{q}$. To handle this issue, we opt for successive convex approximation (SCA) technique, where in each iteration, the right-hand side of (\ref{c2_0}) is replaced by its concave lower bound at a given UAV location denoted by $\textbf{q}^{(l)}$, with $l$ designating the $l^{th}$ iteration. Recalling that any convex
function is globally lower-bounded by its first-order Taylor
expansion at any point, and by following a similar approach as in \cite{Wu2017}, (P2) can be approximated in iteration $l$ by}

{
\begin{subequations}
\small
	\begin{align}
	\max_{\mathbf{q},\boldsymbol{\eta}} & \quad {
	\sum_{k=1}^K w_k \eta_k } \tag{P3} 
	\label{c3_}\\
	\label{c3_0} & \eta_k \leq R_{k}^{k,(l)}, \; \forall k \in \{1,\ldots,K \} \tag{P3.a}\\
	\label{c3_1} & {\text{(\ref{c2_2})}-\text{(\ref{c2_4})}, \tag{P3.b}}
	\end{align}
\end{subequations}
where $R_{k}^{k,(l)}$ is the lower bound of the private signal data rate at the $l^{th}$ iteration, expressed by
\begin{equation}
\small
    \label{eq:LB}
    R_{k}^{k,(l)}=-A_k^{(l)} \left( d_k^{2} - \left(d_k^{(l)}\right)^{2} \right) + B_k^{(l)},
\end{equation}
with
\begin{equation}
\small
    \label{eq:Akl}
    A_{k}^{(l)}=\frac{\log_2(e) |\mathbbm{1} \mathbf{p}_k|^2 }{\left( \sum \limits_{\substack{i=1\\i \neq k}}^K |\mathbbm{1} \mathbf{p}_i|^2 + \sigma^2 \left( d_k^{(l)} \right)^{2}\right) \left( \sum \limits_{i=1}^K |\mathbbm{1} \mathbf{p}_i|^2 +  \sigma^2 \left( d_k^{(l)} \right)^{2} \right) },
\end{equation}
\begin{equation}
\small
    \label{eq:Bkl}
    B_{k}^{(l)}=\log_2\left( 1+{|\mathbbm{1} \mathbf{p}_k|^2}/\left({\sum \limits_{\substack{i=1 \\ i \neq k}}^K |\mathbbm{1}\mathbf{p}_i|^2 + \sigma^2 \left( d_k^{(l)} \right)^{2}}\right) \right),
\end{equation}
and $d_k^{(l)}=||\mathbf{q}^{(l)}-\mathbf{q}_k||_2$, $\forall k=1,\ldots,K$.
Hence, problem (P3) is a convex quadratically constrained quadratic programming problem, which can be solved efficiently using existing software tools such as CVX \cite{CVX}.
}

\subsection{RSMA {Precoding} and Rate-Splitting}
Given {a} UAV location $\textbf{q}$, 
%in addition to $\textbf{w}$ and $\zeta$, 
problem (P1) can be reduced to  

\begin{subequations}
\small
	\begin{align}
	\small
	\max_{\mathbf{P}, \mathbf{r}} & \quad 
	{\text{WSR}=R_{\rm{ov}}} \tag{P4} \\
	\label{c_P31}
	\text{s.t.}\quad & \text{(\ref{c1})--(\ref{c4})}.  \nonumber \tag{P4.a}
	\end{align}
\end{subequations}
Similar to (P1), problem {(P4)} is also non-convex. To solve it, we adopt the same weighted minimum mean square error (WMMSE) approach as in {\cite{Mao2018,Joudeh2016}}, {where (P4) is transformed into an augmented weighted mean square error (AWMSE) problem (P5) as follows.}
{First, any user $k$ ($k=1,\ldots,K$) detects and estimates $s_{0}$ as $\hat{s}_{0}=e_k^{0}y_k$, where $e_k^{0}$ is the equalizer. After successfully decoding $s_{0}$ and subtracting it from the received signal, $s_k$ can be detected and estimated as
\begin{equation}
\small
\hat{s}_k=e_k^k (y_k-\textbf{h}_k^H \textbf{p}_{0}s_{0})=e_k^k (y_k-d_k^{-1} \mathbbm{1}  \textbf{p}_{0}s_{0}).    
\end{equation}
The mean square error (MSE) of each stream can be defined as $\varepsilon_k=\mathbb{E}\{|s_k-\hat{s}_k|^2\}$, calculated as
\begin{equation}
\small
\label{eq:equalizer}
\varepsilon_k^{0}=|e_k^{0}|^2 T_k^{0}- 2 \Re(e_k^{0} d_k^{-1} \mathbbm{1} \textbf{p}_{0})+1
\end{equation}
and
\begin{equation}
\small
\label{eq:equalizer2}
\varepsilon_k^{k}=|e_k^{k}|^2 T_k^{k}- 2 \Re(e_k^{k} d_k^{-1} \mathbbm{1} \textbf{p}_{u})+1,
\end{equation}
where 
%\begin{equation}
$T_k^{0}=\sum_{i=0}^K d_k^{-2}|\mathbbm{1} \textbf{p}_i|^2+\sigma^2$   
%\end{equation}
and 
%\begin{equation}
$T_k^k=T_k^{0}-|d_k^{-1} \mathbbm{1} \textbf{p}_{0}|^2+\sigma^2$    
%\end{equation}
are the received power at user $k$ ($k=1,\ldots,K$) to decode signals $s_{0}$ and $s_k$, respectively. The optimal MMSE equalizers can then be written as \cite{Joudeh2016}
\begin{equation}
\small
\label{eq:equalizeropt}
(e_k^{0})^{\rm MMSE}= \textbf{p}_{0}^H \textbf{h}_k (T_k^{0})^{-1}\;\text{and}\;(e_k^{k})^{\rm MMSE}= \textbf{p}_{k}^H \textbf{h}_k (T_k^{k})^{-1}.
\end{equation}
%and
%\begin{equation}
%\small
%\label{eq:equalizeropt2}
%(e_k^{k})^{\rm MMSE}= \textbf{p}_{k}^H \textbf{h}_k (T_k^{k})^{-1}.
%\end{equation}
By substituting (\ref{eq:equalizeropt}) into (\ref{eq:equalizer})--(\ref{eq:equalizer2}), the MMSEs are written as
\begin{equation}
\small
(\varepsilon_k^{0})^{\rm MMSE}=\min_{f_k^{0}} \varepsilon_k^{0}=(T_k^{0})^{-1} I_k^{0} \; \text{and}\;(\varepsilon_k^{k})^{\rm MMSE}=(T_k^{k})^{-1} I_k^{k},
\end{equation}
%and
%\begin{equation}
%\small
%(\varepsilon_k^{k})^{\rm MMSE}=(T_k^{k})^{-1} I_k^{k},
%\end{equation}
where 
%\begin{equation}
$I_k^{0}=T_k^{0}-d_k^{-2} |\mathbbm{1} \textbf{p}_{0}|^2$
%\end{equation}
and 
%\begin{equation}
$I_k^{k}=T_k^{k}-d_k^{-2}|\mathbbm{1} \textbf{p}_{k}|^2$    
%\end{equation}
are the interference terms when decoding $s_{0}$ and $s_k$, respectively. Subsequently, the SINRs can be expressed by
\begin{equation}
\small
 \gamma_k^{0}=1/((\varepsilon_k^{0})^{\rm MMSE})-1 \; \text{and}\; \gamma_k^{k}=1/((\varepsilon_k^{k})^{\rm MMSE})-1, 
\end{equation}
%and 
%\begin{equation}
%\small
% \gamma_k^{k}=1/((\varepsilon_k^{k})^{\rm MMSE})-1, 
%\end{equation}
and the common and private data rates by
\begin{equation}
\small
 R_k^{0}=-\log_2\left( (\varepsilon_k^{0})^{\rm MMSE}\right) \;\text{and}\;   R_k^{k}=-\log_2\left( (\varepsilon_k^{k})^{\rm MMSE}\right),  
\end{equation}
%and 
%\begin{equation}
%\small
% R_k^{k}=-\log_2\left( (\varepsilon_k^{k})^{\rm MMSE}\right),   
%\end{equation}
respectively. Consequently, the AWMSEs are given by
\begin{equation}
\small
    \label{eq:awmse}
    \zeta_k^{0}= u_k^{0} \varepsilon_k^{0} -\log_2\left( u_k^{0} \right)\; \text{and}\;     \zeta_k^{k}= u_k^{k} \varepsilon_k^{k} -\log_2\left( u_k^{k} \right),
    \end{equation}
 %   and 
 %   \begin{equation}
 %   \small
 %   \label{eq:awmse2}
%    \zeta_k^{k}= u_k^{k} \varepsilon_k^{k} -\log_2\left( u_k^{k} \right),
%\end{equation}
where $u_k^{j}>0$ ($j=0,k$), are weights associated with the MSEs of user $k$ ($k=1,\ldots,K$). By setting the optimization variables as the equalizers and weights, the relation data rates--WMMSEs can be given as \cite{Joudeh2016}
\begin{equation}
\small
\label{eq:wm1}
(\zeta_k^{0})^{\rm MMSE}= \min_{\mu_k^{0},f_k^{0}} \zeta_k^{0} =1-R_k^{0}\;\text{and}\;(\zeta_k^{k})^{\rm MMSE}= \min_{\mu_k^{k},f_k^{k}} \zeta_k^{k}=1-R_k^{k}.
\end{equation}
By substituting (\ref{eq:wm1}) into (\ref{eq:awmse}), and after some manipulations
%\begin{equation}
%\small
%(\zeta_k^{0})^{\rm MMSE}= u_k^{0} (\varepsilon_k^{0})^{\rm MMSE} - \log_2(u_k^{0})\end{equation}
%and
%\begin{equation}
%\small
%(\zeta_k^{k})^{\rm MMSE}= u_k^{k} (\varepsilon_k^{k})^{\rm MMSE} - \log_2(u_k^{k}),
%\end{equation}
%and the optimal MMSEs 
\begin{equation}
\small
\label{eq:mu1}
(u_k^{0})^*=\left((\varepsilon_k^{0})^{\rm MMSE}\right)^{-1}\; \text{and}\;(u_k^{k})^*=\left((\varepsilon_k^{k})^{\rm MMSE}\right)^{-1}.
\end{equation}
}

{Motivated by the data rates-WMMSE relations in (\ref{eq:wm1}), the optimization problem (P4) can be reformulated as
\begin{subequations}
	\begin{align}
	\small
	\min_{\mathbf{P}, \mathbf{v}, \mathbf{u}, \mathbf{e}} & \quad 
	\sum_{u=1}^K w_k \zeta_{k}^{\rm tot} \tag{P5} \\
	\text{s.t.}\quad & 
	\sum_{k=1}^K v_k^0+1 \geq \zeta_{0},  \nonumber \tag{P5.a} \\
	\label{c2_P5} & \zeta_{k}^{\rm tot}\leq 1-R_{k,\rm th},\;\forall k \in \{1,\ldots,K\} \tag{P5.b}\\
	\label{c3_P5} & \text{tr}(\mathbf{P}\mathbf{P}^H)\leq P_t  \tag{P5.c}\\
		\label{c4_P5} & \mathbf{v}\leq \mathbf{0}  \tag{P5.d} 
	\end{align}
\end{subequations}
where $\textbf{v}=[v_1^{0},\ldots,v_K^{0}]=-\textbf{r}$, $\textbf{u}=[u_1^{0},\ldots,u_K^{0},u_1^1,\ldots,u_K^K]$, $\textbf{e}=[e_1^{0},\ldots, e_K^{0}, e_1^1,\ldots,e_K^K]$, $\zeta_k^{\rm tot}=v_k^{0}+\zeta_k^k$ ($k=1,\ldots,K$), and $\zeta_{0}=\max \{\zeta_1^{0},\ldots,\zeta_K^{0}\}$.
}
{When minimizing the objective in (P5) for $\textbf{u}$ and $\textbf{e}$ (fixed $\textbf{P}$ and $\textbf{v}$), the optimal MMSE  ($\textbf{u}^{\rm MMSE}$,$\textbf{e}^{\rm MMSE}$) is obtained 
according to (\ref{eq:equalizeropt}) and (\ref{eq:mu1}). The obtained values satisfy the Karush Kuhn Tucker (KKT) optimality conditions in (P5) for $\textbf{P}$. Thus, given (\ref{eq:wm1}) and the common rate transformation $\textbf{v}=-\textbf{r}$, (P5) can be transformed into (P4). Similarly, for any solution ($\textbf{P}^*$,$\textbf{v}^*$,$\textbf{u}^*$,$\textbf{e}^*$) that satisfies the KKT conditions in (P5), the solution ($\textbf{r}^*=-\textbf{v}^*$,$\textbf{P}^*$) satisfies the KKT conditions in (P4). Consequently, (P4) can be transformed into (P5). However, (P5) is also non-convex for joint parameters optimization. To solve it, Mao \textit{et al.} proposed in \cite{Mao2018} to adopt an AO method as follows. In the $l^{th}$ iteration of the AO algorithm, the equalizers and weights are updated using the precoders obtained in the $(l-1)^{th}$ iteration, i.e.,
\[
(\textbf{u},\textbf{e})=(\textbf{u}^{\rm MMSE}(\textbf{P}[l-1]),\textbf{e}^{\rm MMSE}(\textbf{P}[l-1])).
\]
Then, ($\textbf{v}$,$\textbf{P}$) is updated by solving (P5) for ($\mathbf{u}$,$\mathbf{e}$). Hence, ($\textbf{u}$, $\textbf{e}$) and ($\textbf{v}$, $\textbf{P}$) are iteratively updated until convergence of the WSR. The details of the procedure are presented in Algorithm \ref{Algo2}, where $[l]$ is the iteration index, $\mathbf{u}$ is the stream's weight vector, $\mathbf{e}$ is the equalizer vector, $\mathbf{v}$ is the transformation of $\mathbf{r}$, and $\varepsilon \ll 1$ is the convergence condition \cite{Mao2018}.
}

\begin{algorithm}[t]
\small{
\caption{RSMA {Precoding}-Rate Splitting Algorithm}
\label{Algo2}
\begin{algorithmic}[1]
\State {Initialize $l \xleftarrow{}0$, $\mathbf{P}^{(l)}$, ${R}_{\rm{ov}}^{(l)}$}
\Repeat
\State Set $l \xleftarrow{} l+1$; $\mathbf{P}^{(l-1)} \xleftarrow{} \mathbf{P}$
\State Set $\mathbf{u} \xleftarrow{} \mathbf{u}^{\rm{MMSE}}(\mathbf{P}^{(l-1)})$; $\mathbf{e} \xleftarrow{} \mathbf{e}^{\rm{MMSE}}(\mathbf{P}^{(l-1)})$
\State Solve (P5) for updated ($\mathbf{u}, \mathbf{e}$), then update ($\mathbf{P}$, $\mathbf{v}$)
\Until {$|{R}_{\rm{ov}}^{(l)}-{R}_{\rm{ov}}^{(l-1)}|\leq \varepsilon$}
\State Return ($\textbf{P},-\textbf{v}$) and $R_{\rm{ov}}^{(l)}$ \% since $\textbf{v}=-\textbf{r}$ \cite{Mao2018}
\end{algorithmic}}
\end{algorithm}

\subsection{{Joint UAV Placement and RSMA Parameters Optimization}}
{Given the solutions to the independent UAV placement and RSMA precoding-rate splitting sub-problems, we propose here a joint optimization algorithm that solves (P1) iteratively. Indeed, our solution, shown in Algorithm \ref{Algo3}, alternates between solving (P2) for fixed $\textbf{P}$ and $\textbf{R}$ and solving (P4) for a fixed UAV location. This procedure continues until convergence of the WSR, i.e., the difference between WSR performances of the current and previous iterations is below $\varepsilon$.}

\begin{algorithm}[t]
\small{{
\caption{{Joint UAV Placement and RSMA Parameters Optimization Algorithm}}
\label{Algo3}
\begin{algorithmic}[1]
\State {Initialize $l \xleftarrow{} 0$, ($\textbf{q}^{(l)},\textbf{P}^{(l)}, \textbf{u}^{(l)}, \textbf{e}^{(l)}, \textbf{v}^{(l)}, {R}_{\rm ov}^{(l)}$)}
\State Set $l \xleftarrow{} l+1$
\Repeat 
\State \multiline{Given ($\textbf{P}^{(l-1)}, \textbf{u}^{(l-1)}, \textbf{e}^{(l-1)}, \textbf{v}^{(l-1)}$), solve (P3) using the SCA method and get the solution $\textbf{q}^{(l)}$} 
\State \multiline{Given $\textbf{q}^{(l)}$, solve (P4) using the AO method in Algorithm \ref{Algo2} and get the solution ($\textbf{P}^{(l)}, \textbf{u}^{(l)}, \textbf{e}^{(l)}, \textbf{v}^{(l)}, {R}_{\rm ov}^{(l)}$)}
\State Set $l \xleftarrow{} l+1$
\Until {$|{R}_{\rm{ov}}^{(l-1)}-{R}_{\rm{ov}}^{(l-2)}|\leq \varepsilon$}
\State Return $(\textbf{q}^*, \textbf{P}^*, \textbf{r}^*, R_{\rm{ov}}^*)$
\end{algorithmic}}}
\end{algorithm}

\subsection{{Complexity Analysis}}
{In Algorithm \ref{Algo3}, the complexity of solving problem (P1) resides mainly in the complexity of solving problems (P2) and (P4). The UAV placement problem (P2) is solved using the SCA approach. Since we have $(2K+3)$ constraints in (P3), the required number of iterations by SCA is $\mathcal{O}\left( \sqrt{2K+3} \log_2(1/\varepsilon) \right)$, where $\varepsilon$ is the accuracy of SCA \cite{CVX}. At each iteration, (P3) is solved with complexity $\mathcal{O}\left(Y_1^2 Y_2\right)$, where $Y_1=K+3$ and $Y_2=2K+3$ are the number of variables and constraints, respectively \cite{LOBO1998}. Thus, the overall SCA complexity to solve (P2) is $\mathcal{O}\left(K^{1.5}\log_2\left(1/\varepsilon\right)\right)$. As presented in Algorithm \ref{Algo2}, (P5) can be solved using alternating optimization (AO). For each iteration in Algorithm \ref{Algo2}, the complexity is dominated by step 5, which solves (P5) using the interior-point method. Given $(N_t+1)K$ variables, step 5 has complexity $\mathcal{O}\left( (N_t K)^3 \right)$ \cite{Boyd2004}, and subsequently, the complexity of Algorithm \ref{Algo2} is $\mathcal{O}\left(Y_3 K^3 N_t^3  \right)$, where $Y_3$ is the number of iterations in Algorithm \ref{Algo2}.}
{Finally, the complexity of Algorithm \ref{Algo3} that solves problem (P1) is $\mathcal{O}( Y_4 K^{1.5} \log_2(1/\varepsilon)+ Y_4 Y_3 K^3 N_t^3)$, where $Y_4$ is the number of iterations in Algorithm \ref{Algo3}.}

\section{Numerical Results}
We consider an RSMA-based UAV system, where one $N_t$-antenna UAV-BS is deployed to serve randomly located {$K$ users in an area of $300 \times 300$ m$^2$}. 
For the sake of simplicity, we assume that users are on the ground, i.e., $z_k=0$, {$\forall k\in \{1,\ldots,K\}$}, {the noise power $\sigma^2=1$}, 
%\textcolor{blue}{Unless stated otherwise,} 
%Moreover, we set $\beta=2$, while $\alpha_k$ are generated from the range $[0,100]$, $\forall k \in \{1,2,3\}$. 
%\textcolor{blue}{we select $N_t=4$, the weight vector
{$\textbf{w}=[1,\ldots,1]$ (corresponds to calculating the sum rate)}, %the SIC decoding order policy for 2-order streams as $\zeta=\{12 \xrightarrow{} 13 \xrightarrow{} 23 \}$, 
{the} bandwidth $B=20$ MHz, and {the data rate threshold} $R_{k,\rm{th}}=0$ Mbps.
%, \textcolor{blue}{$\forall k\in \{1,\ldots,K\}$}. 

%\textcolor{blue}{We assume the following initialization for the multiple access schemes. In RSMA, the power ratio between the common and private streams is fixed to 0.7 and 0.3, respectively. Subsequently, we initilize the precoders as $\textbf{p}_k=0.3 \; P_t \; \frac{\mathbf{h}_k}{||{\mathbf{h}_k}||}$ and $\textbf{p}_0=0.7 \; P_t \; \textbf{u}_0$, where $\textbf{u}_0$ is the  largest left singular vector of the channel matrix $\textbf{H}=[\textbf{h}_1 \ldots \textbf{h}_K]$. For SDMA, the precoders are designed according to the maximum ratio transmission (MRT) method \cite{Joudeh2016}, whereas for NOMA, the precoder of the first user is initiated as in RSMA. The precoders of the second to the before last user are initiated similarly, but for channel $\textbf{H}'$ from which the channel vectors of the firstly decoded users are removed. Finally, the precoder of the last user is initialized using MRT. }

\begin{figure}[t]
	\centering
	\includegraphics[width=0.78\linewidth]{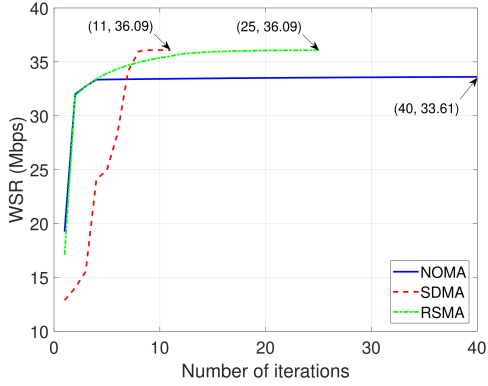}
	\caption{WSR vs. number of iterations ($K=2$, $N_t=2$, SNR$=\frac{P_t}{\sigma^2}=20$ dB, $z_\min=80$ m, $z_\max=120$ m).}
	\label{Fig:00}
\end{figure}
\begin{figure}[t]
	\centering
	\includegraphics[width=0.88\linewidth]{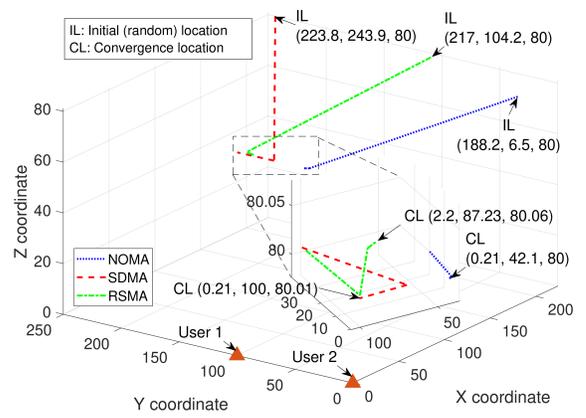}
	\caption{Convergence of the UAV-BS to the best location ($K=2$, $N_t=2$, SNR$=\frac{P_t}{\sigma^2}=20$ dB, $z_\min=80$ m, $z_\max=120$ m).}
	\label{Fig:01}
\end{figure}

{In Figs. \ref{Fig:00}--\ref{Fig:01}, we illustrate the iteration convergence behavior of Algorithm \ref{Algo3} in terms of WSR and UAV-BS location respectively. In Fig. \ref{Fig:00}, we see  that RSMA and SDMA achieve the best performance, with SDMA converging the fastest. Indeed, since the system is underloaded, the number of antennas $N_t$ is sufficient to efficiently handle the multi-user interference. RSMA converges slower than SDMA since it requires more time to adapt its behavior to act like it. Finally, NOMA presents the worst performance since it does not mitigate as efficiently the multi-user interference. 
}

{From Fig. \ref{Fig:01}, we see that the UAV-BS converges to a different location, depending on the used multiple access technique. 
Indeed, the convergence locations are either on or are very close to the Y-Z plan (where the users are located), and are almost at the same minimum allowed altitude $z_\min$ \cite{cherif2020optimal}.
Nevertheless, RSMA and SDMA favor locations close to one of the users, while the UAV-BS for NOMA is closer to the middle point between them. In fact, RSMA and SDMA look for UAV-BS locations that provide non-degraded channels, whereas NOMA prefers a location with degraded air-to-ground channels to perform best \cite{Mao2018,Chen2016}. 
}

\begin{figure}[t]
	\centering
	\includegraphics[width=0.78\linewidth]{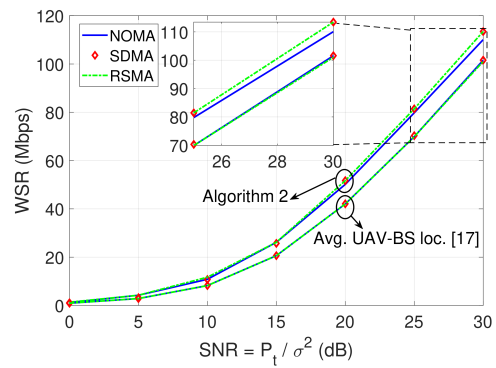}
	\caption{WSR vs. SNR$=\frac{P_t}{\sigma^2}$ ($K=4$, $N_t=4$, $z_\min=80$m, $z_\max=120$m).}
	\label{Fig:03}
\end{figure}

\begin{figure}[t]
	\centering
	\includegraphics[width=0.78\linewidth]{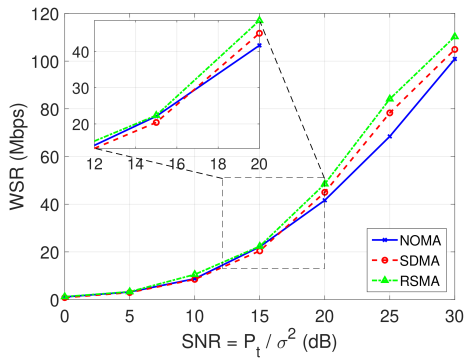}
	\caption{WSR vs. SNR$=\frac{P_t}{\sigma^2}$ ($K=4$, $N_t=4$, $z_\min=80$m, $z_\max=120$m).}
	\label{Fig:02}
\end{figure}

{Fig. \ref{Fig:03} illustrates the WSR performance as a function of SNR for a system of $K=4$ users, with their locations selected as $\textbf{q}_1=[0, 0, 0]$, $\textbf{q}_2=[0, 100, 0]$, $\textbf{q}_3=[150, 150, 0]$, and $\textbf{q}_4=[200, 50, 0]$. The proposed Algorithm \ref{Algo3} is compared to a baseline method, called ``Avg. UAV-BS loc.'', which separates the multiple access problem from the UAV-BS placement one, as in \cite{Liu2019}. Subsequently, we notice that Algorithm \ref{Algo3} outperforms ``Avg. UAV-BS loc.'' for any multiple access scheme. Moreover, RSMA is improved over SDMA and NOMA using our approach, whereas the same performance is achieved by the latter techniques for ``Avg. UAV-BS loc.''. This is due for the optimized UAV-BS locations using Algorithm \ref{Algo3}.}

{Finally, in order to explicitly emphasize the superiority of RSMA over SDMA and NOMA, we adopt for Fig. \ref{Fig:02} the same scenario as Fig. \ref{Fig:03} but with a Rician channel model as follows.}
{The channel coefficient $h_k^j$, ($j=1$ to $N_t$), follows a Rician distribution with parameter $\alpha(\theta_k)$ \cite{Azari2018} and is expressed by
\begin{equation}
\small
    \label{eq:Ricechannel1}
    h_k^{j}={d_k^{-{{\frac{\beta}{2}}}}} {g_k^j(\alpha(\theta_k))}, \qquad \forall j =1,\ldots,N_t,
    %h_k^{j}={d_k^{-{{\frac{\beta(\theta_k)}{2}}}}} {g_k^j(\alpha(\theta_k))}, \qquad \forall j =1,\ldots,N_t,
\end{equation}
where $\theta_k$ is the elevation angle between user $k$ and the UAV, $g_k^j(\alpha(\theta_k))$ is the small-scale fading, which follows a Rician distribution with K-factor $\alpha(\theta_k)$ and $\mathbb{E}\left\{ ||g_k^j||^2 \right\}=1$.}
{Also, $\alpha(\theta_k)=a_1 e^{\theta_k b_1}$, 
%while $\beta(\theta_k)=a_2 Pr_{\rm LoS}(\theta_k)+b_2$ is the path-loss factor, where $Pr_{\rm LoS}(\theta_k)$ is the LoS probability, given by
%\begin{equation}
%\small
%    \label{PLoS1}
%    Pr_{\rm LoS}(\theta_k)=\frac{1}{\left( 1 + a_3 e^{- b_3 (\theta_k- a_3)} \right)},\qquad \forall k=1,\ldots, K,
%\end{equation}
and ($a_1, b_1$)=(10$^{0.5}$, 10$^{1.5}$)
%($a_1, b_1, a_2, b_2, a_3, b_3$)=(10$^{0.5}$, 10$^{1.5}$, -1.5, 3.5, 9.61, 0.16) 
are constants \cite{Azari2018}.}
{For the sake of simplicity, we solve (P4) for the generated Rician channel coefficients, while (P3) is solved based on the large-scale channels only.
%\footnote{\textcolor{blue}{This assumption is motivated by the fact that the presented solution for the UAV-BS location cannot be processed for small-scale varying channels.}}. 
The Rician small-scale coefficients are selected such that the resulting channels are neither aligned, nor orthogonal \cite{Mao2018}.}
{According to Fig. \ref{Fig:02}, RSMA is superior to NOMA and SDMA. Indeed, rate-splitting and precoding, combined to adequate UAV-BS placement, provides non-degraded channels that enable significant WSR gains.
At low SNR, NOMA performs slightly better than SDMA due to the limited power, which performs well over degraded channels. In contrast, SDMA outperforms NOMA at high SNR due to its capability to mitigate interference efficiently.}

\section{Conclusion}
In this paper, we presented {an} in-depth look into the integration of RSMA in UAV{-based} networks. We formulated {the joint UAV placement, beamforming, and rate-splitting} problem, which is non-convex for one UAV-BS. {An alternating optimization } solution was then proposed, where {the UAV placement and the RSMA parameters are optimized iteratively in order to maximize the WSR performance.}
%we divided the problem into two sub-problems.  In the first, the UAV location was optimized to maximize the weighted Rician channel power.  In the second, the joint beamforming and rate-splitting sub-problem was solved to maximize WSR using the WMMSE approach.
The obtained results illustrate the efficiency and robustness of {the proposed approach} compared to other conventional schemes.
%, and the impact of some parameters has been investigated. 
Finally, it is worth noting that {we validate the efficiency of RSMA over NOMA for air-to-ground communications,}
%RSMA outperforms NOMA in different scenarios (terrestrial and aerial networks), %using low-complexity receivers,
which makes it a promising technology for {beyond-5G non-terrestrial networks}.

%\vspace{-13.5pt}

%\appendix[Proof of Lemma \ref{Lemma1}]
%\label{Appendix1}
% or



%\section*{Acknowledgment}

\ifCLASSOPTIONcaptionsoff
  \newpage
\fi

\balance



%\bibliographystyle{IEEEtran}

\bibliographystyle{IEEEtran}
\bibliography{IEEEabrv,tau}

\end{document}